\documentclass[journal]{IEEEtran}

\usepackage{multicol}
\usepackage{booktabs}
\usepackage{graphicx}
\usepackage{subfigure}
\usepackage{float}
\usepackage{standalone}
\usepackage{amsmath}
\usepackage[linesnumbered,lined,ruled,commentsnumbered]{algorithm2e}
\usepackage{comment}
\usepackage{balance}
\usepackage{color}
\usepackage[utf8]{inputenc}
\usepackage{enumitem}
\usepackage{ragged2e}
\usepackage[dvipsnames]{xcolor}
\usepackage{multirow}
\usepackage{graphicx}
\usepackage{cite}
\usepackage{marvosym}

\begin{document}

\title{IoT Platform for COVID-19 Prevention and Control: A Survey}

\author{Yudi~Dong,~\IEEEmembership{Graduate~Student~Member,~IEEE}
        and~Yu-Dong~Yao,~\IEEEmembership{Fellow,~IEEE}
\thanks{Yudi~Dong and Yu-Dong~Yao are with the Electrical and Computer Engineering Department at Stevens Institute of Technology, Hoboken,
NJ, 07030 USA. (E-mail: yyao@stevens.edu)}}

\maketitle


\begin{abstract}
As a result of the worldwide transmission of severe acute respiratory syndrome coronavirus 2 (SARS-CoV-2), coronavirus disease 2019 (COVID-19) has evolved into an unprecedented pandemic.
Currently, with unavailable pharmaceutical treatments and vaccines, this novel coronavirus results in a great impact on public health, human society, and global economy, which is likely to last for many years.
One of the lessons learned from the COVID-19 pandemic is that a long-term system with non-pharmaceutical interventions for preventing and controlling new infectious diseases is desirable to be implemented.
Internet of things (IoT) platform is preferred to be utilized to achieve this goal, due to its ubiquitous sensing ability and seamless connectivity. 
IoT technology is changing our lives through smart healthcare, smart home, and smart city, which aims to build a more convenient and intelligent community. 
This paper presents how the IoT could be incorporated into the epidemic prevention and control system.
Specifically, we demonstrate a potential fog-cloud combined IoT platform that can be used in the systematic and intelligent COVID-19 prevention and control, which involves five interventions including COVID-19 Symptom Diagnosis, Quarantine Monitoring, Contact Tracing \& Social Distancing, COVID-19 Outbreak Forecasting, and SARS-CoV-2 Mutation Tracking.
We investigate and review the state-of-the-art literatures of these five interventions to present the capabilities of IoT in countering against the current COVID-19 pandemic or future infectious disease epidemics.

\end{abstract}


\section{Introduction}

Coronavirus disease 2019 (COVID-19) is a human infectious illness caused by severe acute respiratory syndrome coronavirus 2 (SARS-CoV-2)~\cite{covid19_who}.
SARS-CoV-2 has been spreading all across the world in 213 countries, which results in over 18 million cases of COVID-19 illness as of August 2020~\cite{who_report}. So far, the COVID-19 pandemic situation is not optimistic due to many factors including the unfulfilled vaccine and unavailable pharmaceutical treatment for COVID-19. 
The spreading control of COVID-19 mainly depends on the duration of immunity to SARS-CoV-2 and the non-pharmaceutical interventions (NPIs)~\cite{kissler2020projecting}, such as contact tracing, quarantine, and social distancing. 
On the one hand, the duration of immunity to SARS-CoV-2 is still a mystery, which requires more longitudinal studies to figure out. 
However, it has been known that the immunity to other coronaviruses (e.g, SARS-CoV and MERS-CoV) gradually wanes over time and the coronavirus reinfections exist~\cite{kellam2020dynamics}. Similarly, the immunity to SARS-CoV-2 is highly possible to be short-term instead of permanent. Based on the prediction from Harvard Public Health School~\cite{kissler2020projecting}, the COVID-19 outbreaks would occur recurrently and regularly if the immunity to SARS-CoV-2 is not permanent. 
Even if it is permanent, SARS-CoV-2 could be spreading for many years.

On the other hand, the non-pharmaceutical interventions do contribute to mitigating the severity of COVID-19 epidemics.
A study~\cite{kissler2020projecting} estimates that social distancing could successfully yield a $60$\% reduction of COVID-19 infections~\cite{kissler2020projecting}. 
The combined non-pharmaceutical interventions, involving self-isolation and public events banning, have effectively controlled the transmission of SARS-CoV-2 in some regions like Europe~\cite{flaxman2020estimating} and China~\cite{lai2020effect}. 
Although the non-pharmaceutical interventions are particularly important to COVID-19 prevention and control, it is worth noticing that these interventions probably have profound influence on the economy and society\cite{bonaccorsi2020economic}. Also, the interventions, such as prolonged self-isolation and complete city lockdown, have impacts on people's both physical~\cite{pinto2020combating} and mental health~\cite{fiorillo2020consequences}.
It is thus highly desirable to build a more intelligent and systematic implementation of non-pharmaceutical interventions to ensure effective COVID-19 control with the minimal possible impacts on our lives and society.

\begin{figure*}[t]
  \centering
  \includegraphics[width=\linewidth]{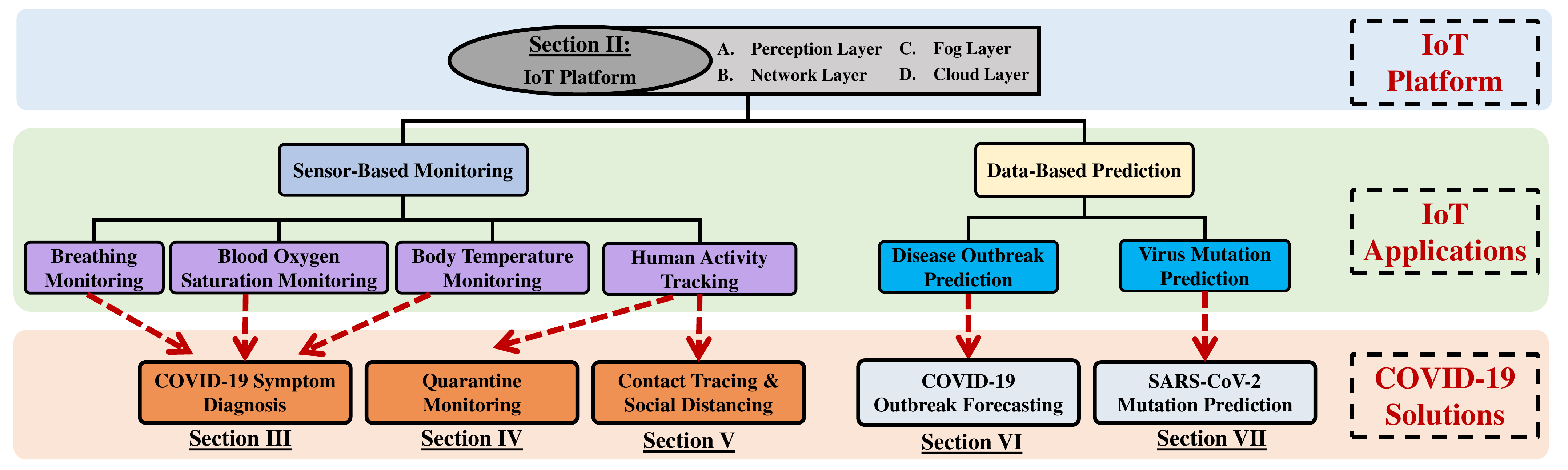}
  \caption{Overview of our survey paper.}
  \label{fig:str}
\end{figure*}

Meanwhile, our living environments are increasingly covered by various sensors inside everyday objects. 
Internet of Things (IoT) technology seamlessly integrates them into the online network and enable them to operate automatically without manual efforts~\cite{li2015internet}. 
It is reported that in 2019, $26.66$ billion IoT devices are utilized and the number will be increased to $35$ billion in 2020~\cite{iotnumb}.
Due to the powerful sensing capability of ubiquitous IoT devices, human features, such as health condition~\cite{yang2014health}, activities~\cite{bianchi2019iot} and vital signs~\cite{haghi2017wearable}, can be captured and analyzed spontaneously in the IoT platform. 
Moreover, the immense data produced by IoT networks can be further explored to perform event prediction using big data analytics and machine/deep learning~\cite{elijah2018overview}.
Thus, with the support of IoT infrastructures accompanied by other emerging technologies (e.g., artificial intelligence (AI), fog computing, and big data), it is feasible to extend the COVID-19 NPIs into our daily lives to achieve intelligent and effective prevention and control.
In this paper, we review an intelligent IoT-based platform for COVID-19 prevention and control that can be used in both the COVID-19 pandemic and post-pandemic periods. 
Specifically, this IoT platform involves three NPIs including \textit{COVID-19 Symptom Diagnosis}, \textit{Quarantine Monitoring}, and \textit{Contact Tracing \& Social Distancing} in a fog layer. 
In a cloud layer, another two NPIs are implemented in the IoT platform, including \textit{COVID-19 Outbreak Forecasting} and \textit{SARS-CoV-2 Mutation Tracking}. 
We comprehensively investigate and review the state-of-the-art studies of IoT-based monitoring and sensing, which can be used to implement these five NPIs for COVID-19 prevention and control. 
Fig.~\ref{fig:str} presents a summary  of our survey paper, which shows the contents of this paper and also illustrates how to associate existing IoT platform and IoT applications with COVID-19 prevention and control. 

The remainder of this paper is organized as follows.
Section~\ref{sec:ecocystem} presents functional layers of a fog-cloud combined IoT platform for COVID-19 prevention and control. We introduce the key techniques in each layer and review the related work in building an IoT platform.
We review the work related to edge-based NPIs in Section~\ref{sec:symptom}, Section~\ref{sec:quarantine}, and Section~\ref{sec:dis}, which refer to \textit{COVID-19 Symptom Diagnosis}, \textit{Quarantine Monitoring}, and \textit{Contact Tracing \& Social Distancing}, respectively. 
In Section~\ref{sec:forcast}, we review the work related to \textit{COVID-19 Outbreak Forecasting}. 
Section~\ref{sec:mutation} investigates the work and future directions in \textit{SARS-CoV-2 Mutation Tracking}.
Finally, we conclude this paper in Section~\ref{sec:con}.

\section{IoT Platform for COVID-19 Prevention and Control: Perception Layer, Network Layer, Fog Layer, Cloud Layer}
\label{sec:ecocystem}
\begin{figure*}[t]
  \centering
  \includegraphics[width=\linewidth]{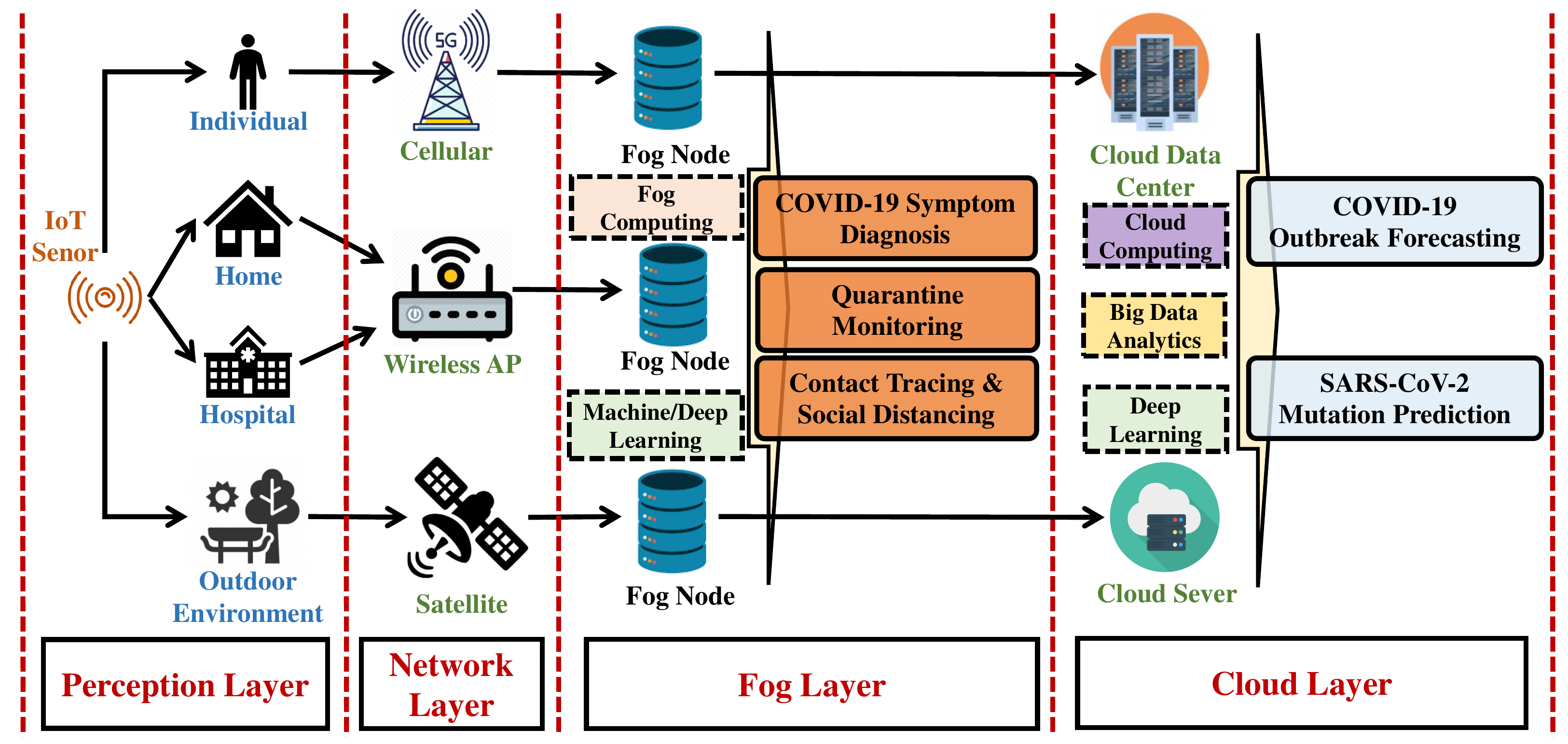}
  \caption{Illustration of an IoT platform for COVID-19 prevention and control.}
  \label{fig:eco}
\end{figure*}

In this section, we introduce a proposed Fog-Cloud-IoT platform for COVID-19 Prevention and Control. 
As illustrated in Fig.~\ref{fig:eco}, we adopt a hierarchical computing architecture, which involves four layers including perception layer, network layer, fog layer, and cloud layer. 
In the first perception layer, it consists of various IoT sensors which are implemented in an individual, home/hospital environment, or outdoor environment to gather all kinds of information, such as vital signs or symptoms of individuals and human activities. 
Next, the sensing data is transmitted through a certain networking technology, such as WiFi, 4G/5G cellular, and satellite. 
Then, distributed fog nodes are deployed with network connections. The fog nodes can be some physical devices (e.g., LAN-connected processor) that are capable of computing and are close to the IoT sensors, which minimizes the latency of real-time data analysis. Therefore, in the fog layer, we can implement time-sensitive NPIs including \textit{Quarantine Monitoring} and \textit{Contact Tracing \& Social Distancing}. 
Finally, all the data streams are fed into a centralized cloud server that has more powerful computing capability. In the cloud layer, we can perform the complex event prediction using sophisticated machining/deep learning algorithms and big data analysis, where two NPIs including \textit{COVID-19 Outbreak Forecasting} and \textit{SARS-CoV-2 Mutation Tracking} are implemented. 

\subsection{Perception Layer}
The perception layer involves various devices that sense the surrounding environment and individuals. 
We specifically describe details of data acquisition in the perception layer by using common IoT sensors.
These IoT sensors are frequently investigated in non-clinical healthcare and human activity sensing.

\subsubsection{Camera}
Camera is a very common and key senor in mobile devices and IoT devices. 
By using the camera, image and video data can be captured and analyzed for various applications, such as non-contacting monitoring and recognition of human activities~\cite{vishwakarma2013survey}.

\subsubsection{Inertial Sensor}
Inertial sensors, equipped in the mobile devices and wearable devices, are the sensors based on inertia and relevant measuring principles~\cite{kostasalexis}, including accelerometer and gyroscope.
Accelerometer, is used for measuring acceleration along three axes. 
By deriving the 3-axis acceleration measurements, it is able to detect the dynamic forces of the device, including gravity, vibrations, and movement. 
Gyroscope is another type of inertial sensors, which is conceptually a spinning wheel with the 3-axis of rotation. It can lead to the measurement of orientation and its rate of change, which tells us how much the device is tilted. 
Human behavior attributes can be derived from the readings of inertial sensors.

\subsubsection{Magnetometer}
Magnetometer is another key sensor in IoT devices, which can detect the magnetic fields along three perpendicular axes. It is originally used to detect the orientation of the device. 
Recent researches show that magnetic fields contain spatial information, which can be exploited to estimate distances between devices based magnetic measurements~\cite{pasku2017magnetic}. Therefore, the magnetometers are studied to perform proximity sensing, which usually serves as a proxy for applications of contact tracing and social distancing.

\subsubsection{Microphone}
Microphone basically is an acoustic sensor that detects and measures ambient sound signals. Current IoT devices typically equip the microphone with micro-electro-mechanical systems (MEMS) technology, which offers a small footprint/thickness, a high signal to noise ratio (SNR), and lower power consumption. Many researches utilize microphones to sense ambient sounds for activity recognition~\cite{zhan2010human,sim2015acoustic}. Moreover, microphones are used in conjunction with speakers, where speakers transmit the deigned acoustic signals and the reflected signals are received by microphones used to analyze Doppler shifts for detecting human activities~\cite{gupta2012soundwave}.

\subsubsection{Commodity WiFi}
Researchers use two main measurements of WiFi, received signal strength indicator (RSSI), and channel state information (CSI), to facilitate sensing tasks. 
First, RSSI characterizes the attenuation of WiFi signals during propagation~\cite{yang2013rssi}.
When people have activities in a WiFi environment, they create perturbations of RSSI, which can be used as the fingerprints of different human activities. 
But RSSI is coarse-grained and unstable, which is easily affected by the environment changes. 
Thus, the fine-grained PHY layer channel state information (CSI) of WiFi signals is recently utilized for sensing tasks~\cite{wang2015understanding}.

\subsubsection{mmWave Radar}
mmWave radar is also explored to achieve non-invasive and non-contacting sensing. 
mmWave radar can modulate the transmitting wireless signals to sweep across a certain frequency band (i.e., frequency modulated continuous wave) and then derive the object movements based on the phase information of reflected signals from objects~\cite{mitomo201077}.
The mmWave radar is robust to environmental changes. For example, light and sound cannot affect its sensing performance.
Since it is capable of precisely modulating the transmitting wireless signals, the mmWave radar is increasingly installed in IoT devices to achieve higher accuracy for activity recognition~\cite{wang2014hybrid}.

\subsubsection{Radio Frequency Identification (RFID)}
RFID is a wireless communications technology that allows recognition of a specific target by radio signals. RFID consists of three components that are a reader, an antenna, and a tag. 
The antenna is connected to the reader and is used to transmit radio signals between the tag and the reader. 
The reader is used to read and recognize the signals from the tag.
The tag is composed of a coupling element, a chip, and a tiny antenna, which is used to receive the radio signals from the reader and antenna. Each tag has a unique electronic code inside and is attached to the object to identify the target object.
RFID is widely applied in healthcare~\cite{amendola2014rfid} and human activity recognition~\cite{ buettner2009recognizing}.

\subsection{Network Layer}
The network layer is responsible for transferring information data or instructions in the perception layer to the whole IoT platform. 
The information transmission relies on the public or private network with the wireless or wired communications mode, which includes 4G/5G cellar networks, WiFi networks, and satellite networks.

\subsection{Fog Layer}
Fog computing is a promising technology introduced by Cisco~\cite{bonomi2012fog}, which is closer to the physical IoT sensors at the network edge comparing with cloud computing and thus brings lower latency for data processing. In the fog layer, the data of IoT devices would be transferred into the corresponding fog node for real-time analysis. 
The fog node can be the devices that are capable of computing, storage, and network connectivity, such as embedded servers or routers. 
Fog nodes are not powerful servers, but a set of low-end and decentralized devices with various functionalities, which is able to infer its own location and track underlying IoT devices to support mobility. 
Due to the low latency and mobility support of fog computing, we can implement time-sensitive and location-sensitive NPIs in the fog layer, which include \textit{COVID-19 Symptom Diagnosis}, \textit{Quarantine Monitoring}, and \textit{Contact Tracing \& Social Distancing}.

\subsection{Cloud Layer}
In the cloud layer, there is a centralized server or data center, which possesses strong information processing and storage capability. The cloud layer is responsible for taking over the tasks that the fog layer is incapable of handling, for instance, the task of complex event prediction. Due to its powerful computing capability, sophisticated algorithms, such as big data analysis algorithms and deep learning algorithms, can be adopted in the cloud layer to improve the system performance. Therefore, another two NPIs including \textit{COVID-19 Outbreak Forecasting} and \textit{SARS-CoV-2 Mutation Tracking} are implemented in this layer.


\section{COVID-19 Symptom Diagnosis}
\label{sec:symptom}
Symptom diagnosis is important for contagion prevention and control. 
By recognizing subjects with symptoms, corresponding interventions (e.g., self-isolation or medication) can be adopted in time, which prevents further transmission of diseases. 
In addition, since hospital visiting is inconvenient and has risks of infection during an outbreak of the pandemic, remote healthcare is desirable. IoT technologies can achieve symptom diagnosis in non-clinical settings and share data with doctors, which makes remote healthcare possible. 
In this section, we review existing IoT-based studies, including breathing monitoring, blood oxygen saturation monitoring, and body temperature monitoring, which can be used for COVID-19 symptom diagnosis. 

\subsection{Breathing Monitoring}
Breathing rates and patterns can reflect the physical condition of individuals, where abnormal breathing patterns could indicate more serious conditions of COVID-19 patients.
Breathing monitoring is thus very important in clinical applications. 
Traditional breathing measurements require hospital visits and professional medical devices attached to the human body, which is not convenient for individuals in need. 
With the advancement of IoT technologies, breathing monitoring are becoming pervasive and ubiquitous. 
Many studies use various IoT sensors, such as inertial sensor~\cite{hernandez2015biowatch,hao2017mindfulwatch,sun2017sleepmonitor}, camera~\cite{murthy2004touchless,abbas2011neonatal,fei2009thermistor,tan2010real,bartula2013camera,nam2016monitoring,massaroni2018contactless,wang2020abnormal}, microphone~\cite{nam2015estimation, martin2017ear, li2017design, faezipour2020smartphone}, mmWave radar~\cite{petkie2009millimeter,lai2010wireless,wang2014application,alizadeh2019remote}, and WiFi~\cite{abdelnasser2015ubibreathe,liu2015tracking,wang2017tensorbeat,wang2017phasebeat}, to continuously monitor breathing activities in both indoor and outdoor environments.

Hernandez~\textit{et al.}~\cite{hernandez2015biowatch} propose a system named BioWatch, which uses the accelerometer and gyroscope of wrist-mounted devices (e.g, smartwatch) to measure both heart rates and breathing rates.
BioWatch exploits a bandpass digit filter with fixed cut-off frequencies of $0.13$ Hz and $0.66$ Hz to extract the breathing signal from the wrist motion data measured by inertial sensors. 
Tian~\textit{et al.}~\cite{hao2017mindfulwatch} further improve the robustness and accuracy of extracting breathing patterns from inertial sensor readings, which utilizes a self-adaptive algorithms to recognize changes in both wrist postures and breathing patterns. 
Moreover, a accelerometer-based sleep monitor~\cite{sun2017sleepmonitor} is proposed to estimate the breathing rates during sleep. It uses a more advanced fusion technique (i.e., Kalman filter) to adaptively merge 3-axes acceleration data to obtain more accurate breathing estimations.  

To achieve non-contacting measurements, Murthy~\textit{et al.}\cite{murthy2004touchless} propose a system that can measure the breathing rates by using thermal camera to capture exhaled air flows near the nose.
More studies~\cite{abbas2011neonatal,fei2009thermistor} then improve the thermal camera-based breathing monitoring system by developing algorithms for automatically tracking the changes near nasal region, which can not only accurately estimate breathing rates but also generate the real-time breathing waveform. 
In addition, Tan~\textit{et al.}~\cite{tan2010real} utilize a regular camera to realize the breathing monitoring by analyzing the chest movements in a camera recorded video. 
Bartula~\textit{et al.}~\cite{bartula2013camera} propose a new algorithm to efficiently distinguish between breathing and non-breathing motions from video streams, which improves the accuracy of breathing monitoring using cameras.
By using the off-the-shelf cameras in mobile devices, Nam~\textit{et al.}~\cite{nam2016monitoring} and Massaroni~\textit{et al.}~\cite{massaroni2018contactless} respectively develop smartphone-based and laptop-based breathing monitoring system to constantly record breathing information. 
Recently, Wang~\textit{et al.}~\cite{wang2020abnormal} propose to use deep neural networks to model the breathing data from depth cameras to recognize the six types of COVID-19 breathing patterns like Eupnea and Tachypnea. 
  
\begin{figure}[t]
  \centering
  \includegraphics[width=0.9\linewidth]{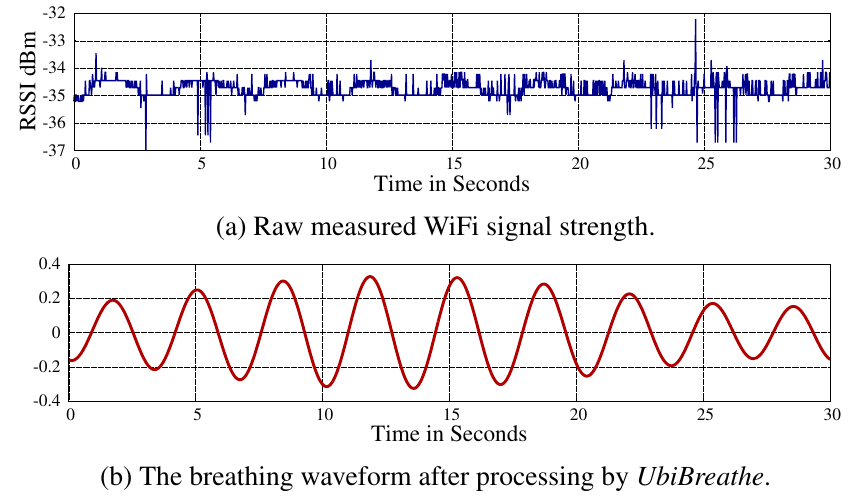}
  \caption{The waveform of WiFi signals while a person is breathing~\cite{abdelnasser2015ubibreathe}.}
  \label{fig:UbiBreathe}
\end{figure}

Although camera-based approaches can achieve non-contacting breathing monitoring, they are constrained by the lighting conditions. Therefore, some studies~\cite{nam2015estimation, martin2017ear, li2017design} explore breathing patterns from breathing sounds recorded by microphones. 
For example, the microphone embedded in smartphones or headsets ~\cite{nam2015estimation} is used to capture the breathing sounds when users place the microphone around the head. 
Similarly, Martin~\textit{et al.}~\cite{martin2017ear} propose to use in-ear microphone to recover the breathing waveform from breathing sounds and also apply a normalized least mean squared adaptive filter to eliminate ambient noises. 
With regard to diagnosing COVID-19 breathing patterns,
Faezipour~\textit{et al.}~\cite{faezipour2020smartphone} use breathing sounds acquired from smartphones to distinguish healthy and unhealthy users based on machine learning models.

Recently, more studies focus on using radio frequency (RF) sensing techniques to monitor breathing motions based on radar and WiFi. 
Petkie~\textit{et al.}~\cite{petkie2009millimeter} use a  continuous-wave (CW) based Doppler radars to measure the chest displacements of breathing based on the Doppler shifted reflected signals. 
To improve the rang resolution of Doppler radar, Lai~\textit{et al.}~\cite{lai2010wireless} propose to use an ultra-wideband (UWB) radar to detect breathing motions, which can achieve high-range resolution for tracking breathing of multiple subjects. But the transmission power of UWB radar is limited, which reduces the signal-to-noise ratio and the sensing range. 
To avoid the disadvantages of the Doppler radar and UWB radar, 
recent studies ~\cite{wang2014application,alizadeh2019remote} utilize frequency modulated continuous wave (FMCW) radars to perform reliable and multi-subject breathing monitoring. 
In addition, the chest and abdominal movements of breathing can impact the WiFi signal indoor propagation (e.g., reflection and scattering), thus it is feasible to capture human breathing motions by analyzing the received WiFi signals. 
Fig.~\ref{fig:UbiBreathe} shows the WiFi signal data of person breathing which indicates clear breathing cycles.  
Abdelnasser~\textit{et al.}~\cite{abdelnasser2015ubibreathe} first propose a WiFi-based breathing estimator named UbiBreathe, which can derive the breathing rates from WiFi signal strength values. 
Liu~\textit{et al.}~\cite{liu2015tracking} utilize the amplitude changes of WiFi channel state information (CSI) to estimate the breathing rates, which is a fine-grained measurement for breathing. 
Further studies~\cite{wang2017tensorbeat}~\cite{wang2017phasebeat} exploit phase information of WiFi CSI to realize accurate breathing monitoring for multi-subjects.


\begin{figure}[t]
  \centering
  \includegraphics[width=0.78\linewidth]{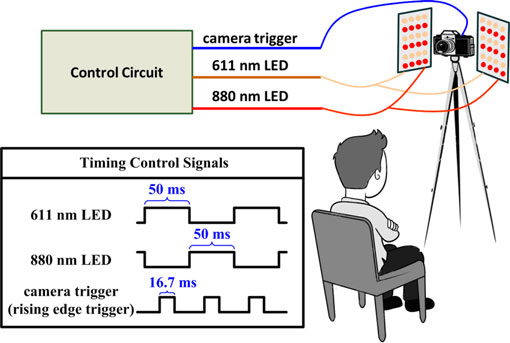}
  \caption{Illustration of camera-based SpO2 measurements~\cite{shao2015noncontact}.}
  \label{fig:cameraspo2}
\end{figure}

\subsection{Blood Oxygen Saturation Monitoring}
Blood oxygen saturation (SpO2) is a measure of the oxygen-carrying capability of red blood cells. 
Healthy individuals can regulate blood oxygen to a high level (i.e., $>95\%$).
For COVID-19 patients, low blood SpO2 is an early warning sign of the disease, where further treatments (e.g., supplemental oxygen) are needed if the blood SpO2 of a COVID-19 patient is lower than $90\%$~\cite{xie2020association}.
In addition, research indicates that many COVID-19 patients are measured as low blood SpO2 even they do not have any discomfort or symptoms~\cite{spo22020}. 
Therefore, monitoring relative changes of blood SpO2 is significant for COVID-19 diagnosis and treatment. 
In the clinical situations, a pulse oximeter~\cite{chan2013pulse, adiputra2018internet}, which is a noninvasive device placed over an individual's finger, is used to continuously measure the blood SpO2 of patients. 
Blood with different oxygen levels has differences in the absorption of near-infrared light. 
Thus, the pulse oximeter emits near-infrared light to pass through the blood of the finger to measure the blood SpO2. 
However, this clinical pulse oximeter is not convenient for the normal usages in daily lives. More studies exploit wrist-mounted oximeters~\cite{son2017design, florez2017blooxy} or wrist-mounted photoplethysmogram (PPG) sensors~\cite{bagha2011real, yang2015spo2} to measure the blood SpO2 by analyzing the absorption difference of reflected light on the blood in wrist. These wrist-mounted sensors can be integrated into smartwatch or Fitbit to continuously monitor the relative changes of blood Spo2. 

Recently, remote SpO2 measurements~\cite{kong2013non, guazzi2015non, shao2015noncontact, lamonaca2015blood} draw more and more attention to researchers. Wieringa~\textit{et al.}~\cite{wieringa2005contactless} propose a pioneering idea that uses a camera with three different wavelengths of LED light to measure blood SpO2. But they did not build a real system or give any measurement results.
Kong~\textit{et al.}~\cite{kong2013non} use two cameras under ambient light to detect a narrow level range (i.e., $97\%-99\%$) of blood SpO2.
To measure the blood SpO2 in a wider range varying from $80\%$ to $100\%$ , Guazzi~\textit{et al.}~\cite{guazzi2015non} utilize a RGB camera to detect the blood SpO2 of an individual $1.5$ m away.
Moreover, as shown in Fig.~\ref{fig:cameraspo2}, Shao~\textit{et al.}~\cite{shao2015noncontact} propose a new approach to measure the blood SpO2 by using the camera-recorded videos of a individual’s face area, which is a low-cost method suitable for SpO2 monitoring in home settings. 
\begin{figure}[t]
  \centering
  \includegraphics[width=0.78\linewidth]{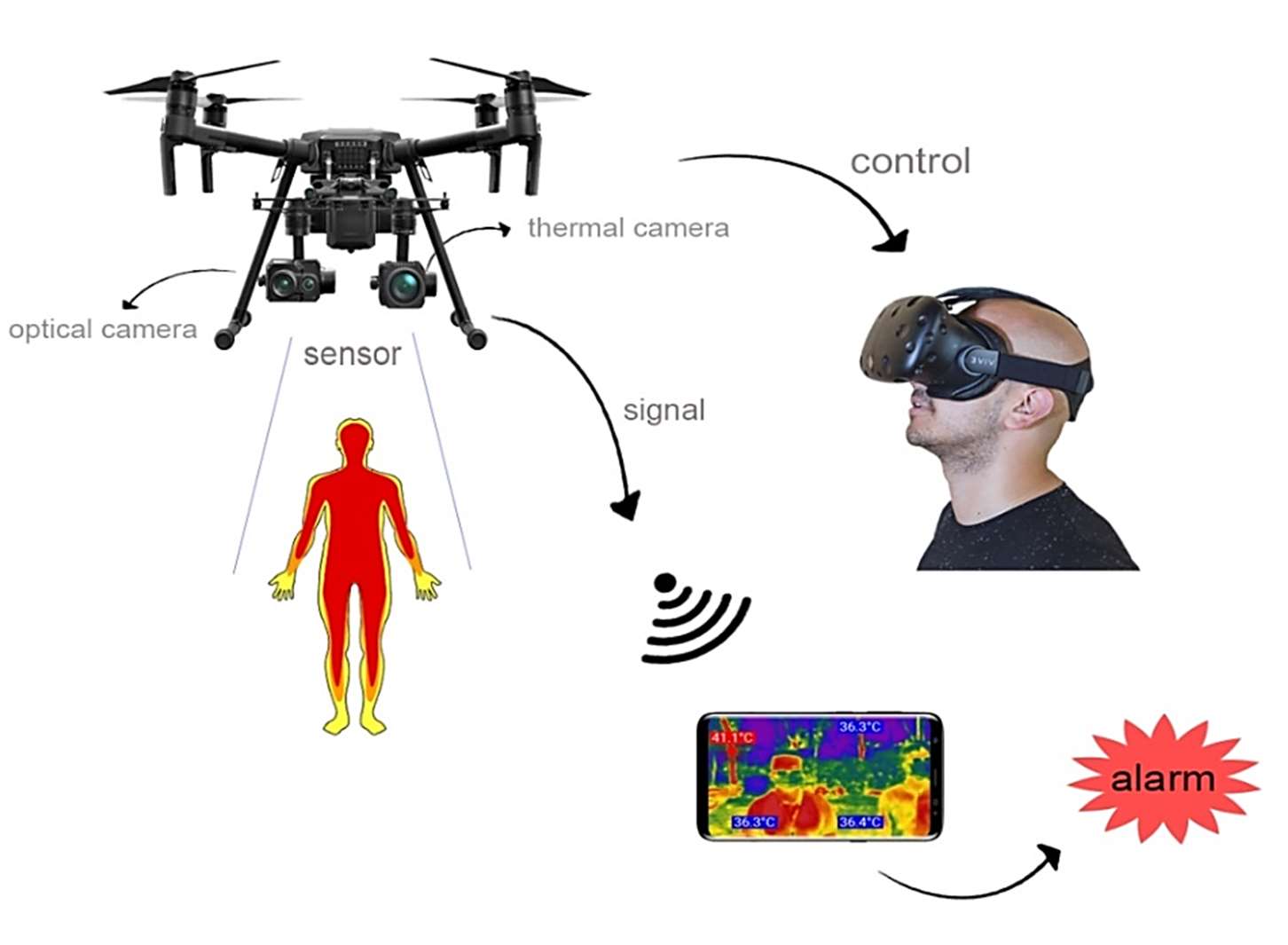}
\caption{Drone-based infrared thermography imaging for COVID-19 detection~\cite{mohammed2020toward}.}
\label{fig:tem}
\end{figure}

\subsection{Body Temperature Monitoring}
Fever is a typical symptom of COVID-19, where clinical statistics shows that more than $80\%$ COVID-19 patients has the symptom of fever~\cite{zhu2020clinical}. 
During the COVID-19 pandemic, many hospitals setup infrared temperature sensors at the entry to identify the febrile patients and isolate them from other patients for further assessment~\cite{hsiao2020measurement}, which significantly reduces the COVID-19 transmission in hospitals.
Thus, it is important for both COVID-19 diagnosis and prevention to monitor the changes of human body temperature. 
Traditional electronic thermometers are based on the principle that different temperatures can cause the changes of probe's resistance and we can measure the resistance of probe to obtain the temperature values.
This thermometer is usually required to be placed in oral, which causes inconvenience and is hard to achieve continuous monitoring. 
Infrared temperature sensors~\cite{ring2012infrared, ng2009remote} thus are commonly used for non-contact measurement of the body temperature in the situation of long-term and continuous monitoring, which are suitable to be implemented in high-risk areas, such as hospitals, schools, and airports, to perform fever screening. For example, Chan~\textit{et al.}~\cite{chan2004screening} explore the feasibility of using an infrared thermography (IRT) camera  to screen fever subjects in the airport during the SARS outbreak. 
The experiments involving $176$ subjects shows that IRT-based approach can recognize fever subjects (i.e., $> 38^{\circ}$C) with accuracy of over $88\%$ in the distance of $0.5$ m. Moreover, many researchers propose IRT-based body temperature monitor for remote healthcare~\cite{zakaria2018iot}~\cite{mandala2017energy}, which enables real-time remote health monitoring.
Similarly, as shown in Fig.~\ref{fig:tem}, Mohammed~\textit{et al.}~\cite{mohammed2020novel}~\cite{mohammed2020toward} propose to use infrared thermography imaging with drones to recognize COVID-19 infected person in the outdoor environment.
However, the accuracy of infrared approaches is limited from a distance. Some studies~\cite{vaz2010full, milici2014epidermal, wen2016wearable} explore RF identification (RFID) technologies for accurate body temperature monitoring. 
They integrate temperature sensors into RFID tags and attach tags onto human skin to enable the body temperature monitoring with an error of $0.25^{\circ}$C.




\section{Quarantine Monitoring}
\label{sec:quarantine}

\begin{figure}[t]
  \centering
  \includegraphics[width=0.9\linewidth]{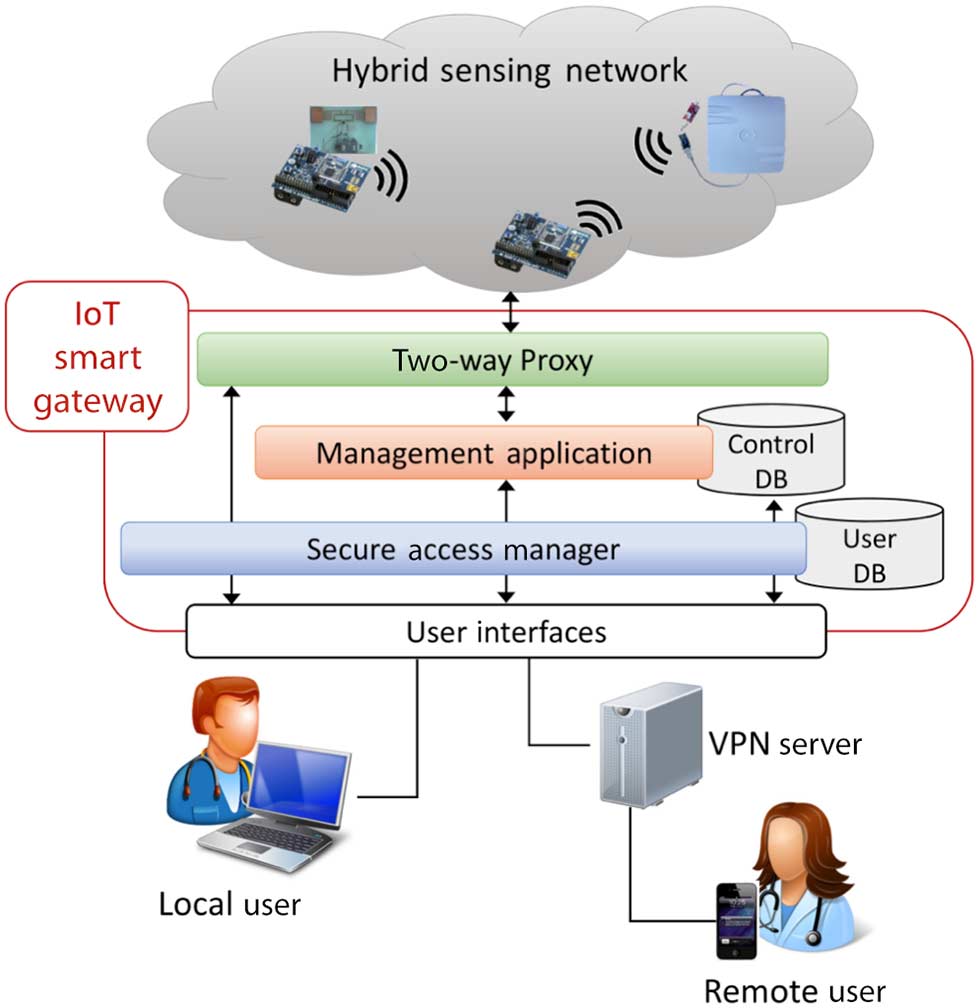}
  \caption{Hybrid sensing network-based IoT health monitoring system~\cite{catarinucci2015iot}.}
  \label{fig:Catarinucci}
\end{figure}

Quarantine is used to isolate the individuals, who have been diagnosed with COVID-19 disease or have been exposed to coronavirus, which is implemented to prevent the spread of COVID-19~\cite{cdcqu}.  
In addition to keeping away from others, their health needs to be monitored for further assessment and possible treatment. 
Conventional procedures of quarantine monitoring, such as vital signs monitoring or activity monitoring, are implemented manually by medical staff.

However, the pandemic brings acute shortages of medical staff and facilities. And this contact monitoring also increases the risk of infections among nursing staff. 
Thus, quarantine monitoring needs to be implemented in home settings during the pandemic.  
The issue is that quarantined individuals at home may not follow the rules and their health cannot be monitored by professional medical staff or devices. 
To address this issue, many studies explore the advances of IoT to enable remote smart healthcare, which achieves automatic human activity tracking and real-time health monitoring in home settings~\cite{baker2017internet}. 
Catarinucci~\textit{et al.}~\cite{catarinucci2015iot} integrate RFID into wireless sensor network (WSN) architecture to build a hybrid sensing network (HSN), which combines the advantages of both RFID and WSN technologies to enable a long-range, low-power consumption and low-cost sensing scheme for patient monitoring. Specifically, the patients need to wear RFID tag for monitoring their physiological data like heartbeat and movement.  
Furthermore, as illustrated in Fig.~\ref{fig:Catarinucci}, the sensing real-time data of HSN is delivered through IoT smart gateway to the local user or remote user (i.e., medical staff) for assessment.
To achieve minimum-latency real-time monitoring, Verma~\textit{et al.}~\cite{verma2018fog} introduce the concept of fog computing to an IoT-based system for remote health monitoring. They integrate a fog layer into an IoT system to process real-time data that delivers the information to doctors timely. 
In addition, a fog-assisted IoT system is proposed in ~\cite{singh2018fog} to monitor the patients of dengue fever, which can remotely send patients' vital signs and symptoms to doctors with short response time.

\begin{figure}[t]
  \centering
  \includegraphics[width=0.9\linewidth]{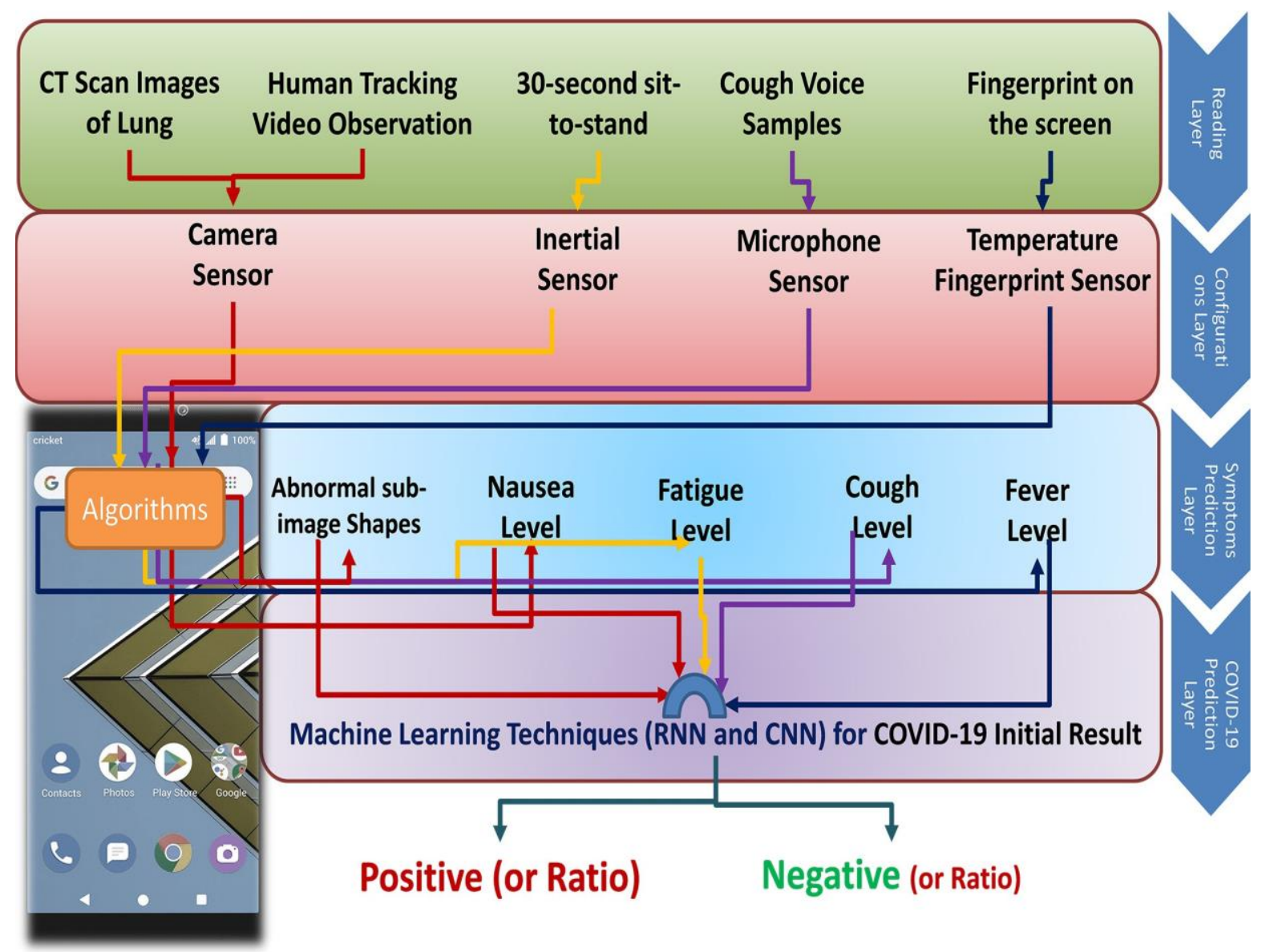}
  \caption{Smartphone-based COVID-19 detection~\cite{maghdid2020novel}.}
  \label{fig:maghdid}
\end{figure}

Many studies~\cite{el2020quarantine,singhiot,otoom2020iot,maghdid2020novel,dobrea2020autonomous} have exploit IoT techniques for the quarantine monitoring of COVID-19 subjects. 
El-Din~\textit{et al.}~\cite{el2020quarantine} propose a basic IoT sensor-based system to monitor COVID-19 infected subjects, which uses ear sensor, blood sensor and motion sensor to measure the patients' physical information (i.e., temperature, respiratory rate, and blood pressure) and sends alerts to hospitals if anomaly is detected. 
Also, Singh~\textit{et al.}~\cite{singhiot} use a wearable band attached to the body to track the real-time locations of COVID-19 quarantine subjects.
Similarly, in ~\cite{otoom2020iot}, an IoT framework is used for monitoring and identify COVID-19 subjects during the quarantine. Specifically, several bio-sensors are deployed to detect the COVID-19 symptoms of subjects and these data would be delivered to quarantine center for further assessment and, in the cloud data center, machine learning algorithms are used to build a model for COVID-19 identification. 

In addition to professional IoT-based bio-sensors, Maghdid~\textit{et al.}~\cite{maghdid2020novel} propose to use the built-in sensors of smartphones to detect the COVID-19 of monitoring subjects at the network's edge. As shown in Fig.~\ref{fig:maghdid}, a smartphone camera and  inertial sensors are used to track the activities of monitoring subjects. The microphone records the acoustic signals of cough and the fingerprint sensor is utilized to measure the body temperature. The sensing data is then fed into a machine learning model to predict COVID-19. This proposed IoT-based framework is a low-cost solution for COVID-19 monitoring. 
Moreover, to achieve an outdoor quarantine monitoring, Dobrea~\textit{et al.}~\cite{dobrea2020autonomous} propose to use a drone with a high-definition camera to monitor the quarantine zones.


\section{Contact Tracing \& Social Distancing}
\label{sec:dis}
Social distancing means keeping a safe distance (i.e.,  $>$ 6 feet or 2 meters) between individuals, which is a very effective intervention for preventing infectious diseases such as coronavirus and influenza virus that spread through droplets while coughing and talking. 
In addition, contact tracing is also a good way to slow down the spread of infectious diseases like COVID-19. Through contact tracing, the close contacts, who did not keep social distancing, are found and required for treatment or self-quarantine to prevent further spread of virus~\cite{contacttracing}. 
Conventionally, those close contacts are provided by the infected individual, which is hard to cover all close contacts and may have omissions. IoT devices can provide a more accurate and convenient way for social distancing and contact tracing.

\begin{figure}[t]
  \centering
  \includegraphics[width=0.9\linewidth]{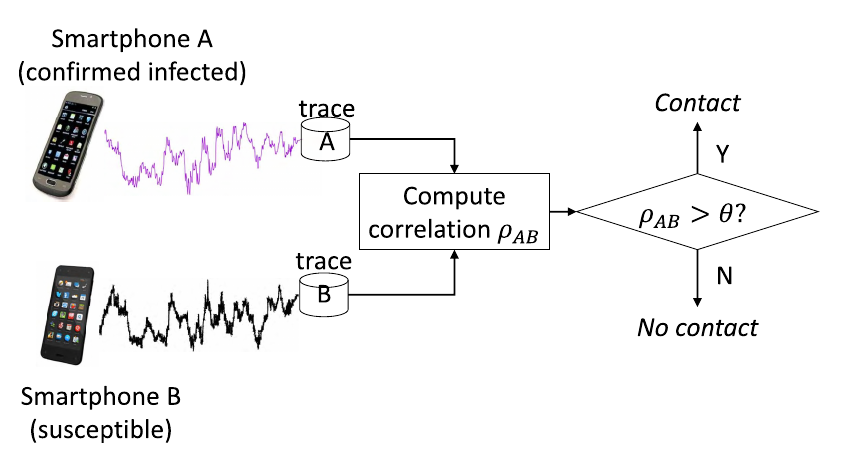}
  \caption{Smartphone magnetometer-based contact tracing and social distancing~\cite{jeong2019smartphone}.}
  \label{fig:jeong}
\end{figure}

IoT devices in possession of various sensors, such as GPS~\cite{paek2010energy}, microphone~\cite{satoh2013ambient,burns2016proximity}, and magnetometer~\cite{jeong2019smartphone}, have been frequently used for proximity detection, which can be implemented for social distancing and contact tracing. 
GPS is a positioning system that gives the coordinates of users. The most intuitive method is using GPS to track the trajectories of users and determine the contact distance based on coordinates~\cite{paek2010energy}. Although the GPS-based method is easy to be realized, it is also obviously flawed due to the high power consumption and low distance resolutions (i.e., 10 meters). Therefore, more researchers explore to integrate high accurate algorithms into power-efficient sensors to achieve proximity detection. 
In~\cite{satoh2013ambient}~\cite{burns2016proximity}, researchers use microphones of smartphones and IoT devices to records ambient sound and then calculate the acoustic power spectrum to estimate distance between users. 
Moreover, researchers find that magnetic fields~\cite{pasku2017magnetic} contain spatial information, which can be explored for proximity detection. 
As shown in Fig.~\ref{fig:jeong}, Jeong~\textit{et al.}~\cite{jeong2019smartphone} propose a magnetometer-based method for contact tracing in epidemics, which exploits linear correlations of smartphone magnetometer readings to estimate distance between two phones to detect the events of close human contact.

\begin{figure}[t]
  \centering
  \includegraphics[width=0.9\linewidth]{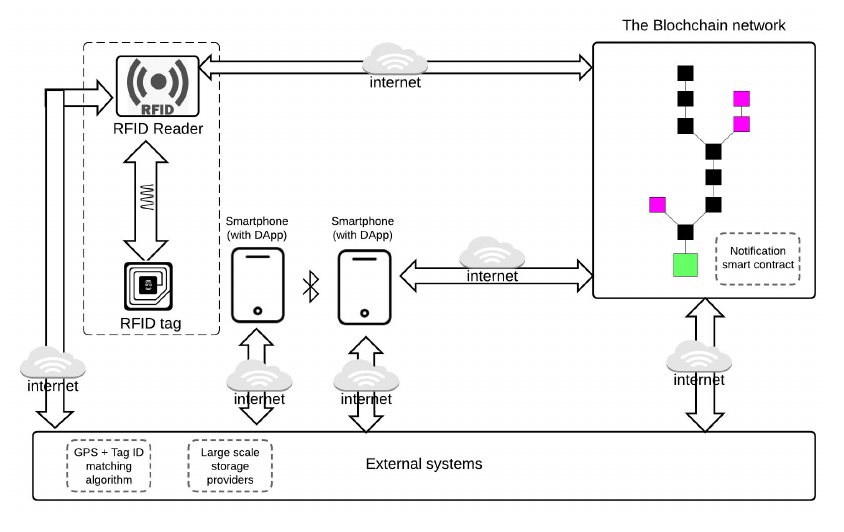}
  \caption{Blockchain diagram for anonymity preserving IoT-based contract tracing~\cite{garg2020anonymity}.}
  \label{fig:garg}
\end{figure}

In addition, RF-based signals, such as Bluetooth~\cite{liu2013face}, WiFi~\cite{sapiezynski2017inferring}, and RFID~\cite{bolic2015proximity}, are explored to detect proximity. 
Liu~\textit{et al.}~\cite{liu2013face} build a model based on Bluetooth signal propagation to map Bluetooth received signal strength values to distance values, which can achieve a precise distance resolution of 1 meter. By using signals from multiple sources (i.e., WiFi and Bluetooth), Sapiezynski~\textit{et al.}~\cite{liu2013face} propose a more accurate and robust system that can estimate distance between individuals with more precise distance resolution of 0.5 meters. 
Bolic~\textit{et al.}~\cite{bolic2015proximity} utilize the backscatter signals from RFID tags to derive the proximity with a small error of 0.3 meters.
Moreover, Farrahi~\textit{et al.}~\cite{farrahi2014epidemic} propose to make cellular communication traces act as a proxy for contact tracing, which use social network information like phone call activities to obtain contact networks of individuals.

Recently, more and more work for COVID-19 contact tracing and social distancing has been reported.
In~\cite{gupta2020enabling}, Gupta~\textit{et al.} envision the smart city and
intelligent transportation system to guarantee social distancing.
Polenta~\textit{et al.}~\cite{polenta2020internet} use WiFi and Bluetooth signals from IoT devices to determine whether two individuals follow social distancing. Also, this work develops a web App for users to manage the collected data. 
Xia~\textit{et al.}~\cite{xia2020return} propose to use Bluetooth Low Energy (BLE) to perform contact tracing based on proximity detection. Also, this work analyzes the relationships between the adoption rate of the contact tracing and COVID-19 control and discusses the security and privacy issues of the contact tracing strategy.
Tedeschi~\textit{et al.}~\cite{tedeschi2020iotrace} propose an IoT-based scheme for COVID-19 contact tracing, named IoTrace. 
IoTrace also uses BLE for distance estimation similar to the previous work. The difference is that IoTrace adopts a decentralized model, which addresses the issues of the location privacy disclosure and the overhead of user devices. As shown in Fig.~\ref{fig:garg}, Garg~\textit{et al.}~\cite{garg2020anonymity} introduce the concept of blockchain to the RFID-based contact tracing, which enhances the security and privacy by using a decentralized IoT architecture.
Moreover, in~\cite{hu2020iot}, authors analyze different architectures of IoT platform that used for COVID-19 contact tracing in terms of protocol stack model and architectural entities.

\section{COVID-19 Outbreak Forecasting}
\label{sec:forcast}

As aforementioned sections present, IoT senors can capture a wealth of data. With the advance of big data analytics and artificial intelligence techniques, we can explore the rich set of IoT underlying data and perform elaborate analysis to predict the occurrence of various events. 
Akbar~\textit{et al.}~\cite{akbar2017predictive} propose a generic architecture for mining IoT data based on machine learning techniques, which can be used for early predictions of complex events. An adaptive prediction algorithms, named adaptive moving window regression, is designed for dynamic IoT data analysis in near real-time. The proposed architecture is implemented in a smart city for predicting traffic events with high accuracy. 
Dami~\textit{et al.}~\cite{dami2018efficient} integrate a latent Dirichlet
allocation model into support vector machine to perform nonlinear data analysis in the IoT environments, which can predict the complex events in an efficient manner with high accuracy.
An IoT-based frost prediction system is proposed by Diedrichs~\textit{et al.}~\cite{diedrichs2018prediction}. This work use IoT sensors of weather stations to capture environmental data like temperature and humidity and exploit Bayesian network and random forest to predict frost events.

\begin{figure}[t]
  \centering
  \includegraphics[width=\linewidth]{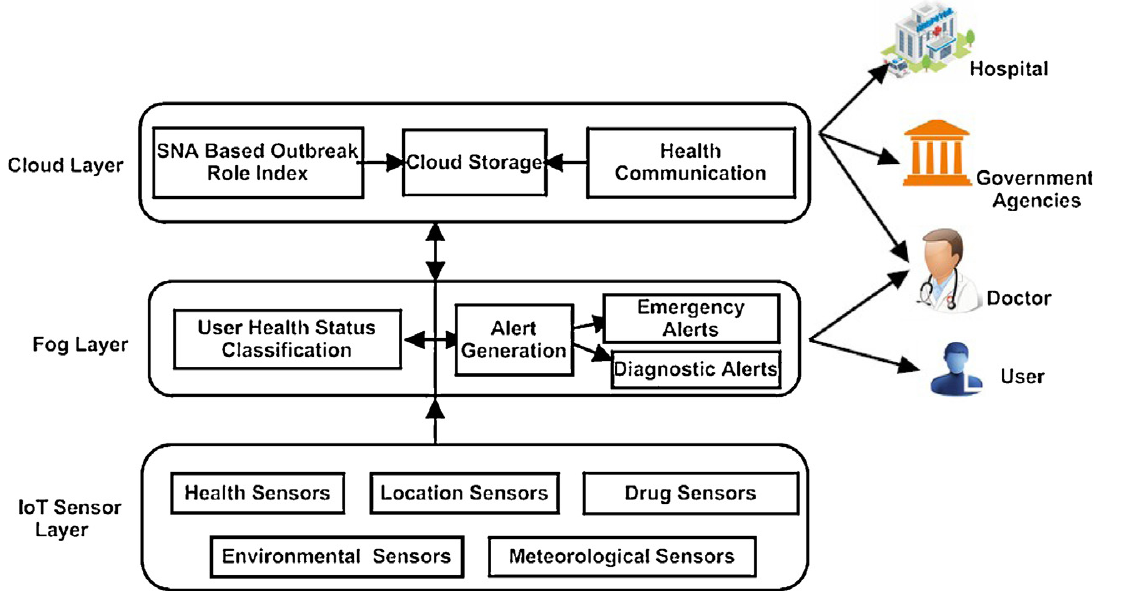}
  \caption{Proposed system~\cite{sood2017wearable} to predict and control outbreaks of chikungunya virus.}
  \label{fig:sood}
\end{figure}

In addition, many researchers have explored IoT data for predicting outbreaks of infectious diseases. 
Sareen~\textit{et al.}~\cite{sareen2017intelligent} design an IoT-based intelligent system to predict Zika virus outbreak. 
Zika virus can cause a mosquito-borne disease. The most common symptom of Zika is fever. This system applies a fuzzy k-nearest neighbour algorithm to recognize the possibly infected users with the fever symptom and uses Google map to locate the infected users for risk assessment. 
For similar arboviruses spreading by mosquitoes, Tavares~\textit{et al.}~\cite{tavares2018iot} propose an IoT architecture for summarizing data from IoT sensors and analyzing them using big data techniques to predict and monitor the arbovirus outbreaks.
As shown in Fig.~\ref{fig:sood}, Sandeep~\textit{et al.}~\cite{sood2017wearable} propose a healthcare system for monitoring and predicting chikungunya virus, which utilizes the advancement of IoT, fog computing, and cloud computing. Chikungunya virus, spreading in many developing countries, can cause vector borne disease. This work~\cite{sood2017wearable} uses IoT-based wearable sensors to acquire the data (i.e., body temperature) of users and implements Fuzzy-C means algorithm in the fog layer to diagnose users in real-time. All the data of infected users is then uploaded to the cloud sever for predicting the outbreak of chikungunya virus using the approach of social network analysis.
Similarly, Rani~\textit{et al.}~\cite{rani2018smart} present an IoT platform for preventing chickungunya virus, where it gathers data from IoT sensors and analyzes them in the cloud using big data processing, and finally gives the suggestions of taking preventive actions. 
Also, for Ebola virus disease, Wesolowski~\textit{et al.}~\cite{wesolowski2014commentary} build a model using mobile network data to analyse the population mobility in Africa that can be useful to forecast and control the Ebola outbreak. 
To summarize, IoT based data analysis has been utilized for predicting various infection diseases. Similar approaches can be implemented for COVID-19 outbreak forecasting.

\begin{table*}
\centering
\small
\caption{Summary of existing IoT-based solutions for COVID-19 prevention and control.}
\label{tab:summary}
\begin{tabular}{|c|c|c|c|c|c|c|} 
\hline
\textit{\textbf{COVID-19 Solutions}}                                                            & \textit{\textbf{IoT Applications}}                                                             & \textit{\textbf{Perception Layer}}                                      & \textit{\textbf{Network Layer}} & \begin{tabular}[c]{@{}c@{}}\textit{\textbf{Fog }}\\\textit{\textbf{Layer}} \end{tabular} & \begin{tabular}[c]{@{}c@{}}\textit{\textbf{Cloud }}\\\textit{\textbf{Layer}} \end{tabular} & \textit{\textbf{Reference}}                                                                                                                                                  \\ 
\hline
\multirow{11}{*}{\begin{tabular}[c]{@{}c@{}}COVID-19 \\ Symptom Diagnosis \end{tabular}}        & \multirow{5}{*}{\begin{tabular}[c]{@{}c@{}}Breathing \\ Monitoring \end{tabular}}              & Inertial sensor                                                         & Cellular                        & $\surd$                                                                                  & -                                                                                          & \cite{hao2017mindfulwatch}                                                                                                                               \\ 
\cline{3-7}
                                                                                                &                                                                                                & Depth camera                                                            & -                               & -                                                                                        & $\surd$                                                                                    &  \cite{wang2020abnormal}                                                                                                                                 \\ 
\cline{3-7}
                                                                                                &                                                                                                & Microphone                                                              & -                               & $\surd$                                                                                  & $\surd$                                                                                    & \cite{faezipour2020smartphone}                                                                                                                           \\ 
\cline{3-7}
                                                                                                &                                                                                                & mmWave radar                                                            & -                               & -                                                                                        & -                                                                                          & \cite{wang2014application}                                                                                                                               \\ 
\cline{3-7}
                                                                                                &                                                                                                & WiFi                                                                    & WiFi                            & $\surd$                                                                                 & -                                                                                          & \cite{abdelnasser2015ubibreathe}                                                                                                                         \\ 
\cline{2-7}
                                                                                                & \multirow{3}{*}{\begin{tabular}[c]{@{}c@{}}Blood Oxygen\\ Saturation Monitoring \end{tabular}} & \begin{tabular}[c]{@{}c@{}}Oximeter\\\end{tabular}                      & -                               & -                                                                                        & -                                                                                          & \cite{son2017design}                                                                                                                                     \\ 
\cline{3-7}
                                                                                                &                                                                                                & PPG sensor                                                              & -                               & -                                                                                        & -                                                                                          & \cite{yang2015spo2}                                                                                                                                      \\ 
\cline{3-7}
                                                                                                &                                                                                                & RGB camera                                                              & -                               & -                                                                                        & -                                                                                          & \cite{shao2015noncontact}                                                                                                                                \\ 
\cline{2-7}
                                                                                                & \multirow{3}{*}{\begin{tabular}[c]{@{}c@{}}Body Temperature\\ Monitoring \end{tabular}}        & Infrared temperature sensor                                             & -                               & -                                                                                        & -                                                                                          & \cite{ring2012infrared}                                                                                                                                  \\ 
\cline{3-7}
                                                                                                &                                                                                                & IRT camera                                                              & Cellular                        & -                                                                                        & $\surd$                                                                                    & \begin{tabular}[c]{@{}c@{}}\cite{zakaria2018iot}\\\cite{mohammed2020novel}\\\cite{mohammed2020toward}\end{tabular}  \\ 
\cline{3-7}
                                                                                                &                                                                                                & RFID                                                                    & -                               & -                                                                                        & -                                                                                          & \cite{wen2016wearable}                                                                                                                                   \\ 
\hline
\multirow{5}{*}{\begin{tabular}[c]{@{}c@{}}Quarantine \\ Monitoring \end{tabular}}              & \multirow{8}{*}{\begin{tabular}[c]{@{}c@{}}Human Activity \\ Tracking \end{tabular}}           & RFID                                                                    & -                               &                                                                                          & $\surd$                                                                                    & \cite{catarinucci2015iot}                                                                                                                                \\ 
\cline{3-7}
                                                                                                &                                                                                                & Smart devices                                                           & WiFi, Cellular                  & $\surd$                                                                                  & $\surd$                                                                                    & \cite{verma2018fog}                                                                                                                                      \\ 
\cline{3-7}
                                                                                                &                                                                                                & \begin{tabular}[c]{@{}c@{}}Ear sensor,\\motion sesnor\end{tabular}      & Cellular                        & -                                                                                        & $\surd$                                                                                    & \cite{el2020quarantine}                                                                                                                                  \\ 
\cline{3-7}
                                                                                                &                                                                                                & Smartphone                                                              & WiFi, Cellular                  & $\surd$                                                                                  & $\surd$                                                                                    & \cite{maghdid2020novel}                                                                                                                                  \\ 
\cline{3-7}
                                                                                                &                                                                                                & Drone, GPS                                                              & Radio~                          & $\surd$                                                                                  & -                                                                                          & \cite{dobrea2020autonomous}                                                                                                                              \\ 
\cline{1-1}\cline{3-7}
\multirow{3}{*}{\begin{tabular}[c]{@{}c@{}} Contact Tracing \\ Social Distancing \end{tabular}} &                                                                                                & WiFi, Bluetooth                                                         & WiFi, Bluetooth                 & -                                                                                        & $\surd$                                                                                    & \cite{polenta2020internet}                                                                                                                               \\ 
\cline{3-7}
                                                                                                &                                                                                                & RFID                                                                    & Cellular                        & $\surd$                                                                                  & -                                                                                          & \cite{garg2020anonymity}                                                                                                                                 \\ 
\cline{3-7}
                                                                                                &                                                                                                & Cellular~trace                                                          & Cellular                        & -                                                                                        & -                                                                                          & \cite{farrahi2014epidemic}                                                                                                                               \\ 
\hline
\multirow{3}{*}{\begin{tabular}[c]{@{}c@{}}COVID-19 \\ Outbreak Forecasting \end{tabular}}      & \multirow{3}{*}{\begin{tabular}[c]{@{}c@{}}Disease Outbreak\\ Prediction \end{tabular}}        & Wearable device                                                         & -                               & $\surd$                                                                                  & $\surd$                                                                                    & \cite{sood2017wearable}                                                                                                                                  \\ 
\cline{3-7}
                                                                                                &                                                                                                & \begin{tabular}[c]{@{}c@{}}Mobile phones \\and body sensor\end{tabular} & Wireless Mode                   & $\surd$                                                                                  & $\surd$                                                                                    & \cite{rani2018smart}                                                                                                                                     \\ 
\cline{3-7}
                                                                                                &                                                                                                & GPS                                                                     & Cellular                        & $\surd$                                                                                  & $\surd$                                                                                    & \cite{wesolowski2014commentary}                                                                                                                          \\ 
\hline
\begin{tabular}[c]{@{}c@{}}SARS-CoV-2 \\ Mutation Prediction \end{tabular}                      & \begin{tabular}[c]{@{}c@{}}Virus Mutation \\ Prediction \end{tabular}                          & -                                                                       & -                               & -                                                                                        & -                                                                                          & -                                                                                                                                                                            \\
\hline
\end{tabular}
\end{table*}
\section{SARS-CoV-2 Mutation Prediction}
\label{sec:mutation}
Similar to most coronavirus, SARS-CoV-2 is a RNA virus with unstable single-stranded structure, which is characterized by a high mutation rate. This mutation may evolve the virus to become more infectious/mortal or drug resistant~\cite{pachetti2020emerging}.
Therefore, tracking and predicting the mutation of SARS-CoV-2 is very significant for proactively preventing and controlling COVID-19.

Traditionally, researchers predict the virus mutation via analyzing the changes of RNA secondary structure~\cite{mathews2010folding, lai2013importance,barash2011mutational}, which involves massive biology laboratory operations. 
With the emerging of AI and big data analysis, more researches~\cite{salama2016prediction,yin2020tempel} use machine/deep learning models to learn the mutation patterns of viral evolution from historical data.
Salama~\textit{et al.}~\cite{salama2016prediction} propose to use a multi-layer perceptron neural network to learn the rules of correlation between nucleotides of RNA. These learned rules can be exploited to predict the mutations in next generations of RNA, where it achieves a prediction accuracy of 75\%. 
Yin~\textit{et al.}~\cite{yin2020tempel} propose a time-series prediction model based on recurrent neural networks (RNN) to perform the mutation prediction of influenza viruses. By leveraging the feature of RNN that can remember all historical residue information, this work improves the effectiveness of the predication model and successfully derive the mutation dynamics of influenza virus.
For SARS-CoV-2, Magar~\textit{et al.}~\cite{magar2020potential} collect 1933 antibody sequences of SARS-CoV-2 and train a model based on these data using support vector machine and multilayer perceptron neural network to predict the possible neutralizing antibodies for SARS-CoV-2. 

The above studies predict virus mutations by exploring the virus structural features. On the other hand, these internal structural features of a virus are always reflected in the external biological features of a virus, such as symptoms of infected subjects, virus transmission rate and virus mortality rate. As we mention in previous sections, all those biological features of a virus could be monitored and analyzed in an IoT platform.
Thus, we consider that exploring gathered data from IoT platforms for COVID-19 applications is also a feasible direction for predicting the mutation patterns of SARS-CoV-2. However, there is no existing studies yet in this direction, and we hope investigations can be explored soon in this important direction.

\section{Conclusions}
\label{sec:con}

This paper proposes a fog-cloud combined IoT platform for COVID-19 prevention and control by implementing five NPIs, including \textit{COVID-19 Symptom Diagnosis}, \textit{Quarantine Monitoring}, \textit{Contact Tracing \& Social Distancing}, \textit{COVID-19 Outbreak Forecasting}, and \textit{SARS-CoV-2 Mutation Tracking}. 
Table~\ref{tab:summary} summarizes various studies considering different layers of the proposed IoT platform. 
We review the recent IoT-based studies which can be applied for implementing the five NPIs. We discuss how the recent technological advancements such as fog computing, clouding computing, artificial intelligence, and big data analysis, can be utilized for IoT and COVID-19 applications. 



%


\bibliographystyle{IEEEtran}
\bibliography{ref}

\end{document}